\documentclass[aps,twocolumn,a4paper,showpacs]{revtex4}
\usepackage{graphicx}
\usepackage{amssymb}
\usepackage{amsmath}
\newcommand{\be}{\begin{equation}}
\newcommand{\ee}{\end{equation}}
\newcommand{\ben}{\begin{eqnarray}}
\newcommand{\een}{\end{eqnarray}}
\newcommand{\bes}{\begin{subequations}}
\newcommand{\ees}{\end{subequations}}
\newcommand{\bb}{\bibitem}

\newcommand{\bfi}{\begin{figure}}
\newcommand{\efi}{\end{figure}}
\newcommand{\bc}{\begin{center}}
\newcommand{\ec}{\end{center}}

\newcommand{\LL}{{\cal L}}
\begin{document}
\title{New results on twinlike models}
\author{D. Bazeia$^{1,2}$ and R. Menezes$^{2,3}$}
\affiliation{$^1$Departamento de F\'{\i}sica, Universidade Federal da Para\'{\i}ba, 58051-970 Jo\~ao Pessoa, PB, Brazil}
\affiliation{$^2$Departamento de F\'{\i}sica, Universidade Federal de Campina Grande, 58109-970 Campina Grande, PB, Brazil}
\affiliation{$^3$Departamento de Ci\^encias Exatas, Universidade Federal da Para\'{\i}ba, 58297-000 Rio Tinto, PB, Brazil}


\begin{abstract}
In this work we study the presence of kinks in models described by a single real scalar field in bidimensional spacetime.
We work within the first-order framework, and we show how to write first-order differential
equations that solve the equations of motion. The first-order equations strongly simplify the study of linear stability,
which is implemented on general grounds. They also lead to a direct investigation of twinlike theories,
which is used to introduce a family of models that support the same defect structure, with the very same energy density and linear stability.  
\end{abstract}

\pacs{11.27.+d, 11.10.Kk}

\maketitle
\section{Introduction}

Kinks and lumps are defect structures of current interest in high energy physics \cite{vs,ms} and in other areas of nonlinear science \cite{wa}.
In high energy physics, in particular, kinks and lumps appear in models described by a single real scalar field in bidimensional spacetime,
and they are static solutions of the corresponding equations of motion. Recent examples of models where kinks 
play important role can be found, for instance, in Refs.~{\cite{tt,dunne,ba,juan,bec1,bec2,mr,exp,rs,as,ak}}. See also Ref.~\cite{lump},
where one studies bell-shaped or lump-like solutions in models described by a single real scalar field.

In Cosmology, defect structures are usually related to the existence of phase transitions in the early Universe. Motivated by Cosmology, however, another direction of study concerns models where kinematic modifications of the scalar field are introduced \cite{c1}, aimed to contribute to explain
the present accelerated expansion of the Universe. Recent investigations of scalar fields with modified kinematics have been presented in \cite{s1,s2,s3,b1,b2,b3,b4,acg} 
with distinct motivations, in particular to search for the presence of topological structures which live in a compact region, and for applications to the braneworld context, with the five dimensional spacetime engendering a single extra dimension of infinite extent. For other recent applications within the brane-world context, see, e.g., Ref.~\cite{B}.

In the study of models with modified kinematics, in \cite{altw} the authors noted that it is possible for k-defects to masquerade as canonical
scalar field solutions. That is, given a standard scalar field model, the topological defect profile and corresponding energy density
can also appear in a k-field theory. In this case the two models are twins of each other, and the idea was further explored very recently in \cite{bdglm}, and also in \cite{aq}.
Motivated by these recent results on twinlike models \cite{altw,bdglm,aq}, in the present paper we introduce and study a new family of models, described by a single real scalar field that can accommodate interesting results. Before doing that, however, in the next Sec.~\ref{rev} we briefly review some results obtained in the case of standard models. We then move on to the main results of the present work, which are described in Sec.~\ref{new}, where we discuss the new family of models, and in Sec.~\ref{twin}, where we study twinlike models. We end the paper in Sec.~\ref{end}, with some comments and conclusions.

\section{Standard models}
\label{rev}

We start with a scalar field $\phi$, described by the standard Lagrange density
\be\label{sm}
\LL= \frac12 \partial_\mu \phi \partial^\mu \phi - U(\phi)
\ee
where $U(\phi)$ is the potential, which is used to identify the particular model under consideration. In this work we deal with bidimensional spacetime, with metric such that $x^0=x_0=t$ and $x^1=-x_1=x$. For simplicity, we use dimensionless field and coordinates, so all the quantities are also dimensionless.

We suppose that the above model engenders static solution, with $\phi=\phi(x)$ that obeys the equation of motion
\be\label{em2}
\phi^{\prime\prime}=U_\phi
\ee
where prime stands for the derivative with respect to $x$, that is, $\phi^\prime=d\phi/dx$, and we are using $U_\phi$ to represent $dU/d\phi$. The energy density of the static solution is given by 
\be\label{ene}
\varepsilon(x)=\frac12 \phi^{\prime 2}+U(\phi)
\ee

We can integrate the equation of motion \eqref{em2} to obtain
\be\label{em1C}
\phi^{\prime2}=2U(\phi)+C
\ee
where $C$ is a constant. If one considers static field with finite energy, we have to take $C=0$. This leads to the equation
\be\label{em1}
\phi^{\prime2}=2U(\phi)
\ee
which is a first-order differential equation that describes static solution with vanishing stress component of the energy-momentum tensor.
We can use \eqref{em1} to introduce an interesting result, concerning the equipartition between the gradient and potential portions of the energy density. We take
\be\label{ene1}
\epsilon_g(x)=\frac12 \phi^{\prime 2}\ee
as the gradient part of the energy density, and
\be
\epsilon_p(x)=U(\phi(x))
\ee
as the potential part. We see that 
\be\label{ene2}
\varepsilon(x)=\varepsilon_g(x)+\varepsilon_p(x)=2\varepsilon_g(x)=2\varepsilon_p(x)
\ee
for the static solution that solves the first-order differential Eq.~\eqref{em1}. Similar expressions also works for the energy, and we can write $E=E_g+E_p=2E_g=2E_p$.

This result is general, and it is valid under the above kinematics, which represents the standard situation, described by the Lagrange density \eqref{sm}. However, if one changes kinematics, the energy equipartition is in general lost. In this work we focus our attention on a new family of models, which represents an interesting possibility to depart from the standard case, since it will lead to new results on twinlike field theories.

\section{New models}\label{new}

Before introducing the family of models, let us introduce the quantity
\be
X=\frac12\partial_\mu\phi\partial^\mu\phi
\ee
Thus, we can write the standard Lagrange density \eqref{sm} in the form
\be
{\cal{L}}=X-U(\phi)
\ee

Usually, one goes beyond the standard case with the two possibilities: we take $F(X)$ as a general function of $X$ and we consider the case
\be
{\cal L}=F(X)-U(\phi)
\ee
or 
\be
{\cal L}=U(\phi)F(X)
\ee
which includes the Born-Infeld generalization.
In this paper, however, we choose a different possibility, described by
\be\label{actionF}
\LL=-U(\phi)F\left(Y\right)
\ee  
where 
\be
Y=-\frac{X}{U(\phi)}=-\frac12\dfrac{\partial_\mu \phi \partial^\mu \phi}{U(\phi)}
\ee
Here we note that to specify the model, we need to have both $U(\phi)$ and the function $F(Y)$, and we see that in the case
\be
F(Y)=1+Y
\ee
we get back to the standard model.
 
The equation of motion for the family of models has the form
\be\label{eof2d}
\partial_\mu \left( F_Y \partial^\mu \phi\right) +  \left(F-Y F_Y\right) U_\phi=0
\ee
where $F_Y=dF/dY$.

The energy-momentum tensor is given by 
\be
T_{\mu\nu}=F_Y\partial_\mu \phi \partial_\nu \phi + g_{\mu\nu} U(\phi) F\left(Y\right)
\ee
or, in components
\bes
\ben
T_{00}&=& F_Y \dot\phi^{2} + U(\phi)F   \\
T_{01}&=&F_Y \phi^\prime \dot \phi \\
T_{11}&=& F_Y \phi^{\prime2} - U(\phi)F \label{stressT11}
\een
\ees
where we are using dot to represent time derivative. Let us now suppose that $\{v_i,\; i=1,2,...,n\}$ is a set of static and uniform solutions of the equation of motion.  This means that $U_\phi(v_i)$ has to vanish. Also, from the energy density we take $U(v_i)=0$ to make the energy of the static and uniform solutions vanish, in the same way they do in the standard case. 

Since we are dealing with a new family of models, we guide  ourselves with the null energy condition (NEC), that is,  we impose  that  $T_{\mu\nu} n^\mu n^\nu\geq 0$, where  $n^\mu$ is a null vector, obeying $g_{\mu\nu}n^\mu n^\nu=0$. This condition  leads to following restriction $F_Y\geq0$, for the general field configuration $\phi(x,t)$ which solves the equation of motion \eqref{eof2d}.

Let us now search for defects structures, considering the case of a static field configuration, $\phi=\phi(x)$. In this case, the equation of motion \eqref{eof2d} changes to
\be
\left(2F_{YY} Y + F_Y \right) \phi^{\prime\prime} = U_\phi \left(F-F_Y Y +2F_{YY} Y^2 \right) 
\ee 
This equation can be integrated to give
\be\label{f_order}
2YF_Y-F=\frac{C}{U(\phi)}
\ee
where $C$ is a constant. In the case of static solution, $Y$ becomes
\be
Y=\frac{\phi^{\prime 2}}{2U(\phi)}
\ee
Thus, the above Eq.~\eqref{f_order} is a first-order equation, and we see that its solutions make $T_{11}=C$.
Thus, the constant of integration that leads us to the first-order equation \eqref{f_order} also identifies the stress component of the energy-momentum tensor.

The first-order equation \eqref{f_order} can be written as follows
 \be\label{forderm111}
 \frac12\phi^{\prime2}=G\left(\frac{C}{U(\phi)}\right)U(\phi)
 \ee
where the function $G$ is such that $G^{-1}(Y)=2YF_Y-F$.
Note that when $C$ vanishes, we get to the case of stressless solutions, and we obtain an algebraic equation for $Y$, such that $F=2YF_Y$; for $G(0)=c$, with $c$ constant, real,
this leads to a simpler relation 
\be\label{Forder}
\phi^{\prime2}=2c \, U(\phi)
\ee
and so we have $Y=c$. Note that the solution $\phi(x)$ of this equation is the solution $\phi_s(x)$ of the equation of motion \eqref{em1} of the standard model, with the position changed as
$x\to\sqrt{c}\;x$, that is, we can write
\be\label{stressless}
\phi(x)=\phi_s(\sqrt{c}\, x)
\ee 
The thickness of the solution is given by 
\be
\delta = \delta_s/\sqrt{c}
\ee
so the solution is thicker or thinner depending on the value of $c$ being lesser or greater than unit. We note that the case of $c$ negative is not allowed; also,
only stressless solutions have the form given by Eq.~\eqref{stressless}.
 
We see that the profile of the stressless solution is similar to the profile of the solution in the standard model. Also, the energy density of the static solutions have the form $T_{00}=U(\phi)F(Y)$.
In the case of stressless solutions we have $C=0$ and we get 
\be\label{FForder}
T_{00}=F(c) U(\phi)
\ee
Here we can write
\be
E=F(c)\int^{\infty}_{-\infty} U(\phi(x)) dx 
\ee
or better
\be\label{relationE}
E= \frac{F(c)}{\sqrt{c}} \int^{\infty}_{-\infty} {U(\phi_s(y))} dy=\frac{F(c)}{2\sqrt{c}}E_s
\ee
where $E_s$ is the energy of the standard solution.

The solutions with a non-vanishing $T_{11}$ are completely distinct from the corresponding solutions of the standard model, since they do not have the form given by Eq.~\eqref{stressless}. Here we recall that for $T_{11}=C$, only the stressless solutions are stable \cite{s2}. Usually, the energy of the other solutions are divergent, and the solutions have oscillatory or divergent profiles as they do, for instance, in the standard model. However, it is interesting to observe that for some choices of the function $F(Y)$ we get to the case of $T_{11}=C\neq0$, with solutions with topological behavior and finite energy. We further illustrate this new possibility with an example below.

\subsection{Linear Stabilty}

Let us now investigate linear stability of the static solution.
We introduce small fluctuations $\eta(x,t)$ about the solution in the usual way: we write $\phi(x,t)=\phi(x)+\eta(x,t)$, where $\phi(x)$ represents the static solution. The  equation of motion \eqref{eof2d} allows obtaining, up to the first-order power in $\eta$,
\ben\label{estabilityeq}
\partial_\mu 
\Big[F_Y \partial^\mu\eta -\frac{F_{YY}}{U(\phi)}\partial^\mu \phi \partial_\alpha \phi \partial^\alpha \eta
\Big]\nonumber
\\ 
=
\Big[U_{\phi\phi}\left(F_YY-F\right)-\frac{U_\phi^2}{U(\phi)}(F_YY-F)_Y Y
\nonumber\\
+\partial_\mu \left(\frac{U_\phi}{U(\phi)}F_{YY} Y \partial^\mu \phi\right)\Big]\eta
\een
Since $\phi=\phi(x)$ is static solution, we take $\eta(t,x)=\eta(x)\cos(\omega\,t)$ to obtain
\ben
-\left[\left(2F_{YY}Y+F_Y\right)\eta^\prime(x)\right]^\prime\nonumber
=
\\ \nonumber
 \Big[U_{\phi\phi} \left( F_YY\!-\!F \right) -\frac{U_\phi^2}{U(\phi)}(F_YY-F)_Y Y\\-\!\Big(\frac{U_\phi F_{YY} Y \phi^\prime}{U(\phi)} \Big)^\prime +\omega^2  F_Y\Big]\!\eta(x)
\een
In the case of the standard model we have $F(Y)=1+Y$, and so we get
\be\label{stability4}
-\eta_s^{\prime\prime}(x)
+ {U_{\phi\phi}}_{|{\phi=\phi_s(x)}} \eta_s(x) = {\omega_s^2}\eta_s(x)
\ee
which describes the standard case, as expected.

In the general situation, for stressless solutions we use the equation \eqref{f_order} with $C=0$ to obtain 
\be\label{stability2}
-\eta^{\prime\prime}(x)
+c\, {U_{\phi\phi}}_{|{\phi=\phi_s(\sqrt{c}x)}} \eta(x) =\frac{\omega^2}{A^2}   \,\eta(x)
\ee
where 
\be\label{hyperboli}
A^2=\frac{2F_{YY}Y+F_{Y}}{F_Y}
\ee
To ensure hyperbolicity of the differential equation, we have to impose the constraint that $A^2$ is constant and positive; see, e.g., Ref.~\cite{s1},
and remember that $A^{-2}$ is the sound speed.

If we introduce the transformation $z=\sqrt{c}\,x$, we can rewrite the above equation \eqref{stability2} in the form
\be\label{stability3}
-\eta^{\prime\prime}(z)
+ {U_{\phi\phi}}_{|{\phi=\phi_s(z)}} \eta(z) = \frac{\omega^2}{A^2\,c} \,\eta(z)
\ee
which has the very same form of the Eq.~\eqref{stability4}, which describes stability in the standard model. Thus, we can find a one-to-one correspondence between the eigenvalues ($\omega$) and eigenstates ($\eta$) of the general theory and the eigenvalues ($\omega_s$) and eigenstates ($\eta_s$) of the standard model. They are related by 
\bes\ben\label{VFVF}
\omega&=&\omega_s A\sqrt{c}
\\
\eta(x)&=&\eta_s(\sqrt{c}\,x)
\een\ees
This is an important result, since the behavior of the stressless solution in the general model given by the Lagrange density \eqref{actionF} is similar to the standard model, differing only by a scaling in $x$ and in the eigenvalues. In the standard model, if we choose a potential $U(\phi)$ that leads to stable static solution, then the general theory also supports stable static solution, under the assumptions that $A^2>0$ and $c>0$.
 
To end the study of stability, let us go further and investigate the second-order contribution in $\eta$ which appear in the energy density.
They are given by
\ben
Tö^{(2)}_{00}\!&=&\!\frac12 \omega^2\eta^2 {F_Y} \sin^2(\omega\,t)\nonumber \\ 
&&+\frac{1}{2} F_Y A^2\left(\eta^{\prime2}+c\,U_{\phi\phi}\eta^2\right)\cos^2(\omega\,t)
\een
Since the contribution is time-dependent, we take the average in time to get 
\be
<\!T^{(2)}_{00}\!>=\frac{F_Y}{4} \left[A^2\eta^{\prime2}+(c\,A^2U_{\phi\phi}+\omega^2)\eta^2\right]
\ee
Now, using the Eq.~\eqref{stability2} we obtain
\be
\!<\!T^{(2)}_{00}\!>=\frac
{F_Y}2 \omega^2\eta^2
\ee
We integrate this expression to get
\ben
\!<\!E^{(2)}\!>\!&=&F_Y \omega^2\int_{-\infty}^{\infty} \eta^2(x) dx \nonumber \\&=&F_Y \omega_s^2 A^2 \sqrt{c}\int_{-\infty}^{\infty} \eta_s^2(z) dz\nonumber \\
&=&\frac{F(c)}{2\sqrt{c}} A^2 \!<\!E^{(2)}_{s}\!>\!
\een
Using the relation \eqref{eof2d},  we can write
\be\label{FFer}
\frac{<\!E^{(2)}\!>}{<\!E^{(2)}_{s}\!>} =\frac{E}{E_s}\,A^2
\ee
which shows that for an admissible $F(Y)$, it is easier or harder to perturb the solution, depending on $A$ being lesser or greater then unit, respectively.   

\subsection{Examples}

Let us now specify the function $F(Y)$ in order to illustrate how the formalism introduced above works. First we recall that for $
F(Y)=1+Y$, we get back to the standard model. 

We introduce a model that can be understood as an extension of the standard model where the dynamics engenders a nontrivial factor, of the form ${\cal L}=g(\phi,X) X - U(\phi)$. In particular, we choose the following Lagrange density  
\be
{\cal L}=\frac{\beta}{2^n} \partial_\mu \phi \partial^\mu\phi\left|\frac{\partial_\mu \phi \partial^\mu\phi}{U(\phi)}\right|^{n-1} \!- \alpha U(\phi)
\ee
Here we suppose that $n\geq1$ is real parameter, and $\alpha$ and $\beta$ are real and positive. Note that for $n=1$, the model describes standard kinematics. This model is different from the model introduced in Ref. \cite{s3} and also investigated in Ref. \cite{b2,b3}. It can be written in the form given by Eq.~\eqref{actionF}, if we identify $F(Y)$ as
\be\label{FY1}
F(Y)=\alpha+\beta Y|Y|^{n-1}
\ee
In this case, both the NEC and hyperbolicity conditions are satisfied. Using the first-order Eq.~\eqref{f_order}, we get
\be\label{phiphi}
\phi^{\prime2}\!=2\!\left|\!\frac{1}{\beta(2n-1)}\!\left({C\,U(\phi)^{n-1}}+\alpha U(\phi)^{n}\right)\!\right|^{\frac{1}{n}}
\ee
We note that for $n>1$, we can find topological solutions even though we consider $C>0$. Such solutions have the same kink profile of the stressless solution, but now they are thinner than the stressless solution. This behavior reflects the fact that solutions with $C$ greater then zero must have energy greater then the stressless solution, so they must be thinner than the stressless solution. In Fig.~1 we depict the above expression \eqref{phiphi} in the $(\phi,\phi^\prime)$ plane for the standard potential
\be
U(\phi)=\frac12(1-\phi^2)^2
\ee

\begin{figure}[t]
\includegraphics[width=3.2cm]{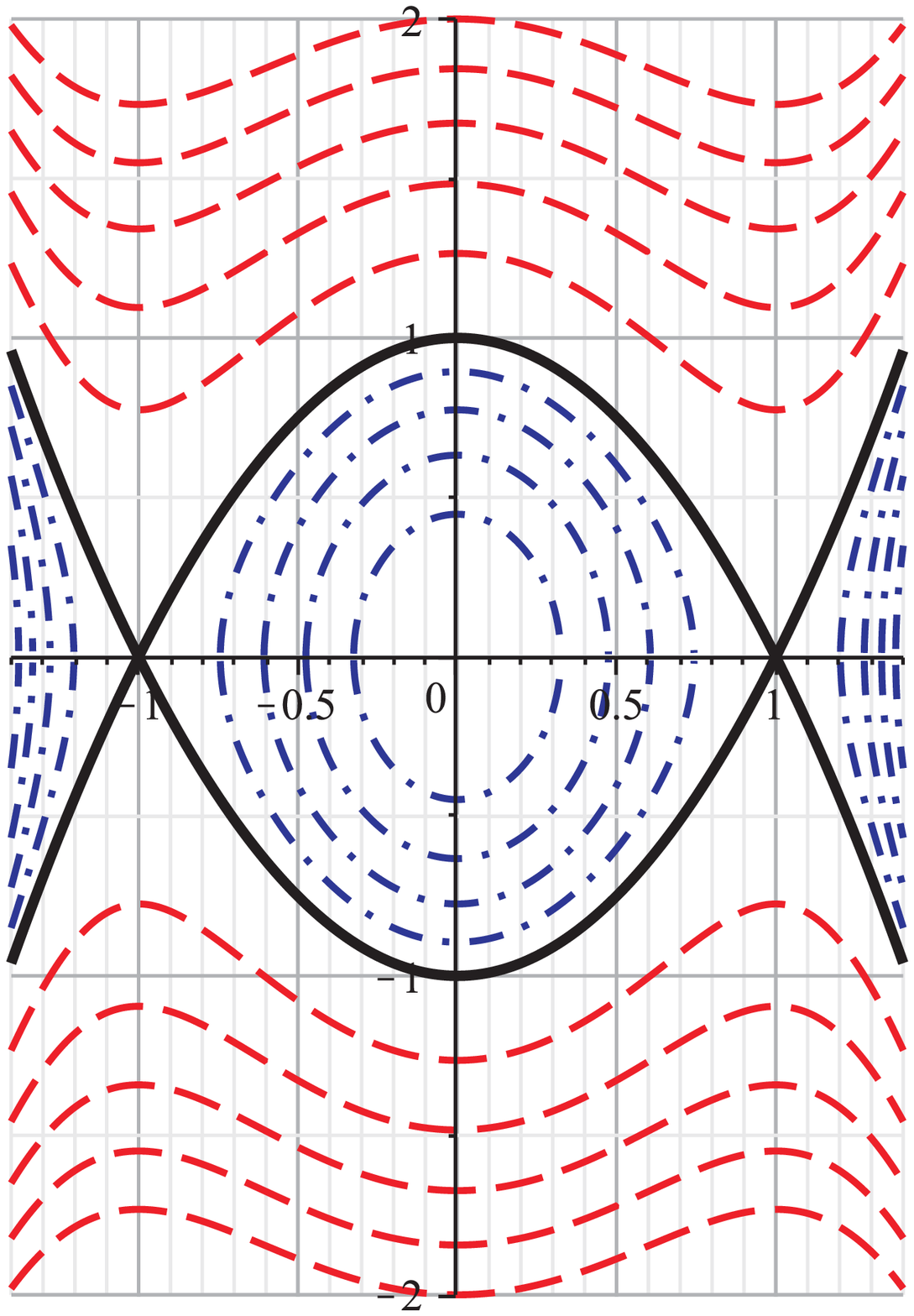}\hspace{0.8cm}
\includegraphics[width=3.2cm]{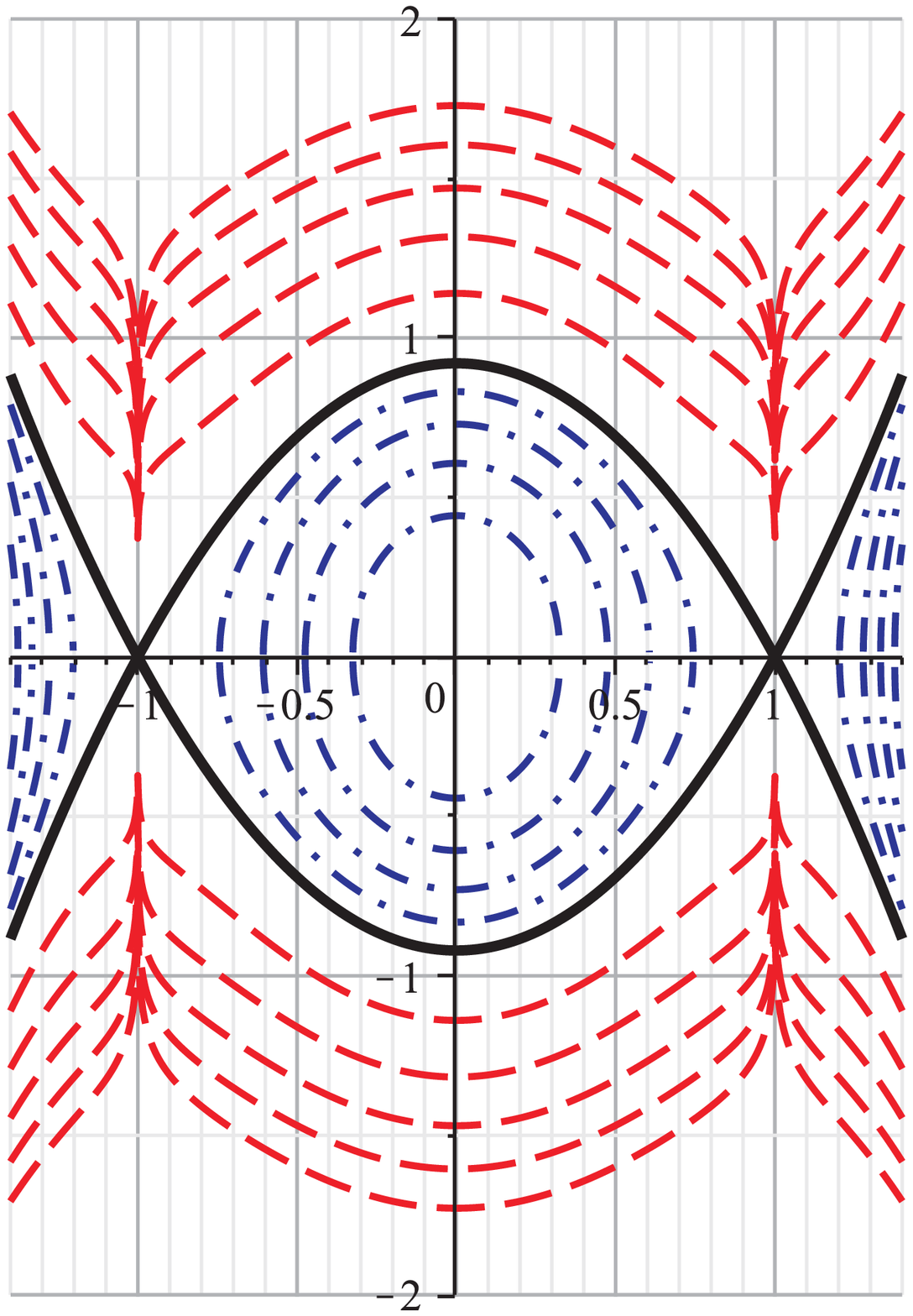}\vspace{0.08cm}
\includegraphics[width=3.2cm]{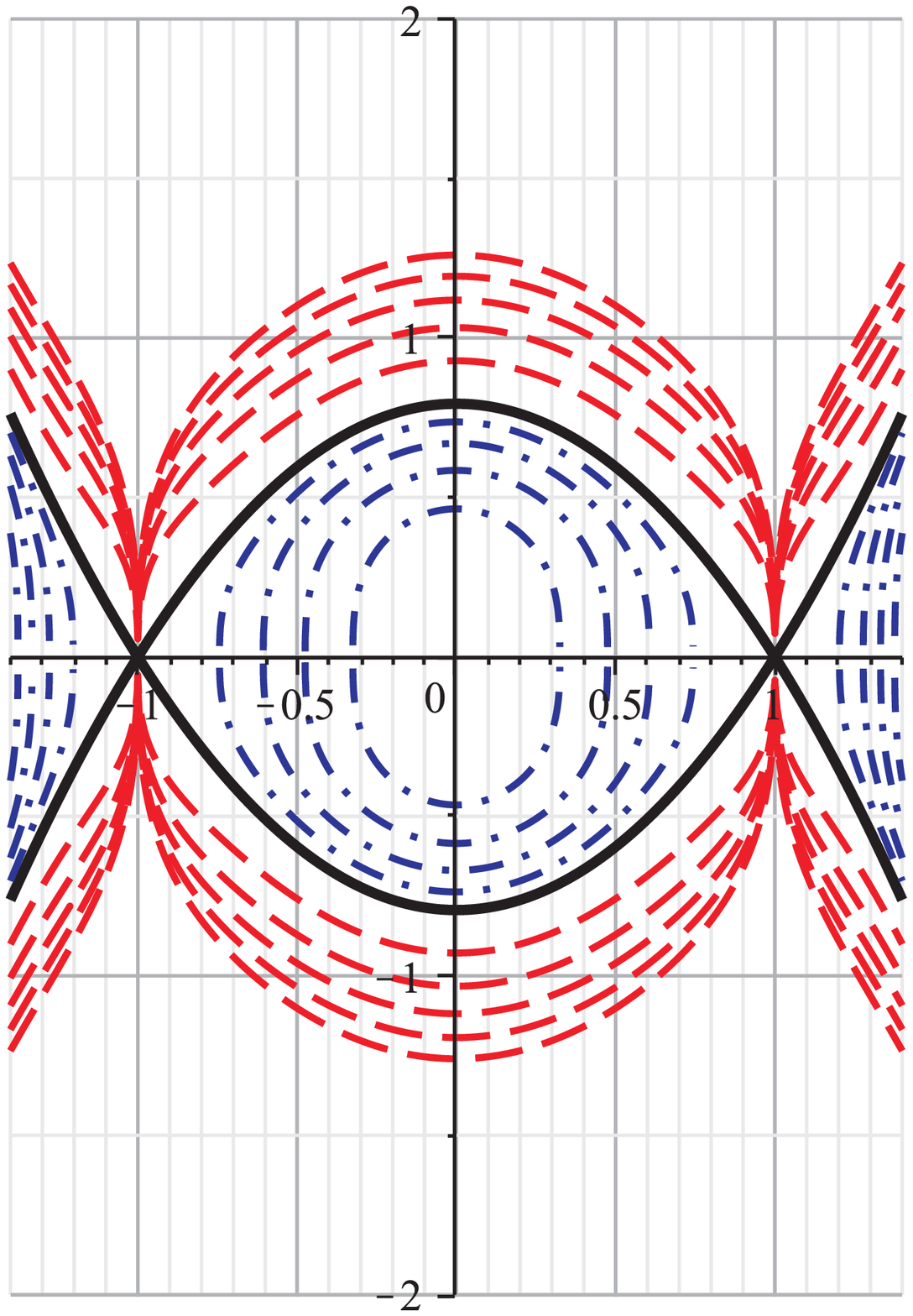}\hspace{0.8cm}
\includegraphics[width=3.2cm]{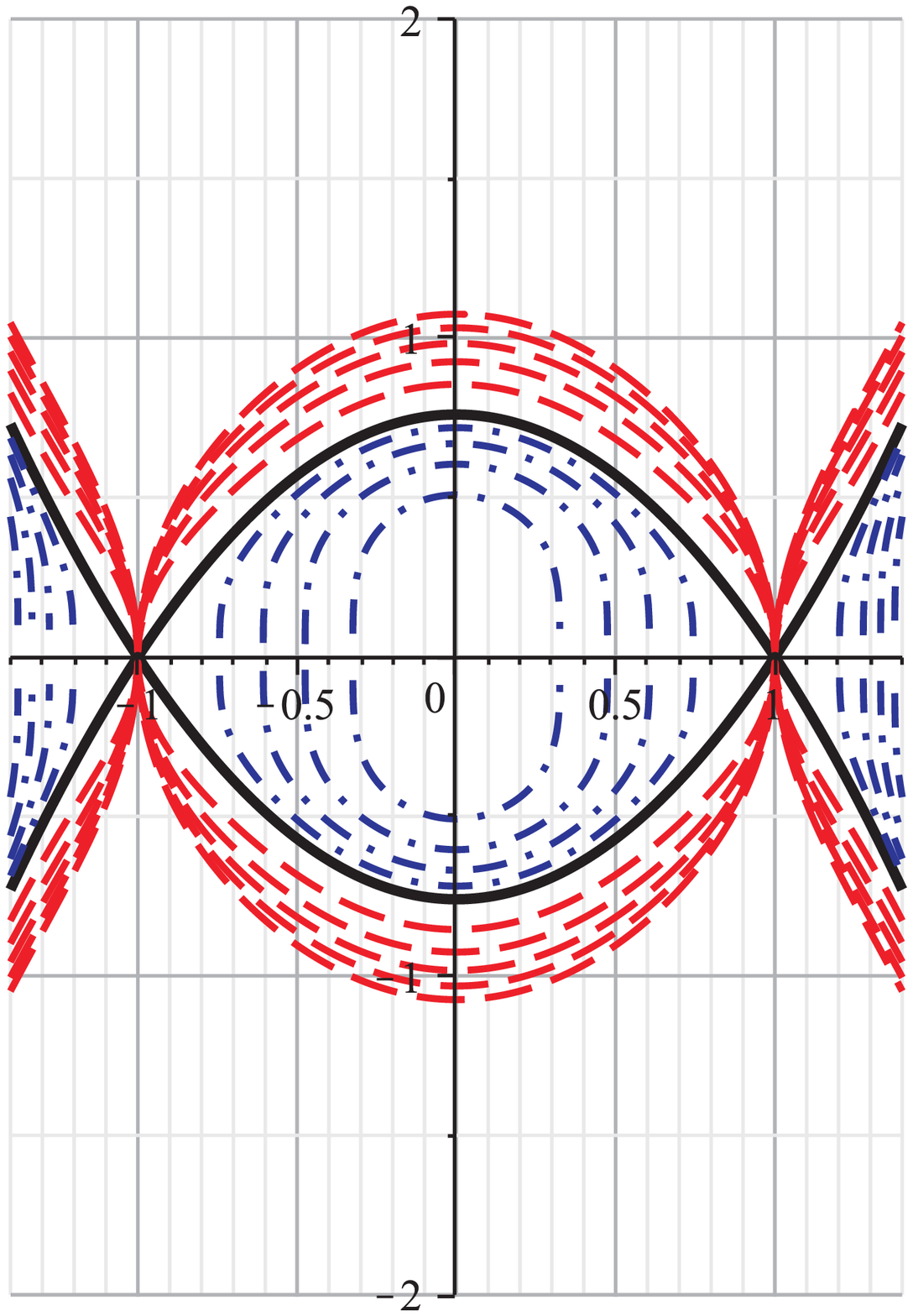}
\caption{Plot of the solutions in the $(\phi,\,\phi^\prime)$ plane, for $C=0$ (black, solid line), $C<0$ (blue, dashed-dotted line) and $C>0$ (red, dashed line), from left to right and up to down, for $n=1,\,1.1,\,1.5,\,$ and $2$, and for $\alpha=\beta=1$.}
\end{figure}

We take $C=0$ in the above Eq.~\eqref{phiphi} to consider the stressless solution. We get
\be
c=\left|\frac{\alpha}{\beta(2n-1)}\right|^{\frac{1}{n}} 
\ee
and so
\be
\frac{E}{E_s}=\left|\frac{\alpha}{2n-1}\right|^{\frac{2n-1}{2n}}\frac{n}{\beta}
\ee
We can fix the thickness of the defect as in the standard model \eqref{sm}, choosing $c=1$; this leads us to the choice $\alpha=(2n-1)\beta$. The energy behaves as
\be
\frac{E}{E_s}=\frac{\alpha\,n}{2n-1} 
\ee
and so it decreases with increasing $n$. The stability of the solution does not depend on $\alpha$ and $\beta$, and we get
\be
A^2=2n-1
\ee
Using the relation \eqref{FFer}, we see that when $n$ increases, it becomes harder and harder to destabilize the solution.

We introduce another model, described by 
\be\label{Taon}
\LL=-U(\phi)\sqrt{1-\frac{\partial_\mu \phi \partial^\mu \phi}{U(\phi)}} 
\ee
If we make the transformation $\partial_\mu \phi\partial^\mu \phi={U(\phi(T))}\,\partial_\mu T\partial^\mu T$, we get $\LL=-U(\phi(T))\sqrt{1-\partial_\mu T \partial^\mu T}$
which can be used to describe the dynamics of tackyonic brane; see, e.g., Ref.~\cite{sen}.

The Lagrange density \eqref{Taon} has the form of the general model \eqref{actionF}, where the $F(Y)$ function is 
\be\label{ftak}
F(Y)=\sqrt{1+2Y}
\ee
The condition \eqref{forderm111} gives
\be
\phi^{\prime2}=\frac{U^2(\phi)}{C^2}-U(\phi)
\ee
The model has no solutions when $C$ is nonzero. For a stressless configuration, $C=0$,  we see that $Y\to\infty$. The solution has the step-like behavior
\be\label{sol2}
\phi(x)=v_1 +(v_2-v_1) H(x)
\ee 
where $v_1$ and $v_2$ are minima states, and $H(x)$ is the Heaviside function. It has the profile of a topological structure with vanishing thickness.
The energy density corresponding to \eqref{sol2} has the form $T_{00}=(\sqrt{2}/2) \delta(x-x_0)$, where $x_0$ is the center of the solution, and 
the energy is finite, given by $E=\sqrt{2}/2$.

In this case we can calculate that 
\be\label{aTaq}
A^2=\frac{1}{1+2Y}
\ee
which vanishes for $Y\to \infty$, that is, for the stressless solution, with $C=0$. This is similar to what happens in the cuscuton model \cite{acg}, the difference is that here this happens with the stressless solution, while in the cuscuton model every solution presents this behavior. 

In order to regularize the solution we follow the case considered in \cite{bmr} and we add a constant to $F(Y)$, that is, we rewrite \eqref{ftak} as
\be\label{FTqReg}
F(Y)=b\left(\sqrt{1+2Y}-\frac1a\right)
\ee
where $a$ and $b$ are real parameters. The solution now obeys 
\be
Y=\frac{a^2-1}{2}
\ee
and the model supports valid solutions for $a>1$, with energy 
\be
E=\frac{b}{\sqrt{2}a}\sqrt{a^2-1}
\ee
We note that $A^2$ obeys the same expression given by Eq. \eqref{aTaq}; therefore, $A^2=a^{-2}$, and $a$ is the sound speed.

\section{Twinlike Models}
\label{twin}

According to some recent results \cite{altw,bdglm,aq}, two distinct models $\LL_1$ and $\LL_2$ are twinlike models if they support static solutions with the same profile, and with the very same energy density.

The family of models here introduced is controlled by $F(Y)$ and $U(\phi)$. Thus, let us now introduce conditions on the function $F(Y)$ for a theory given by the action \eqref{actionF} to be twin of the standard theory, where $F(Y)=1+Y$. Comparing the former Eqs.~\eqref{em1} and \eqref{Forder}, for $c=1$ we see that they have the same static solution, and this fulfills one of the two condictions for the models to be twins.
In this case, the value of $Y$ at the static solution has to equal unit, that is, we have to impose that $Y=1$ for the static solution. Next, we compare Eqs.~\eqref{ene1},\,\eqref{ene2}, and \eqref{FForder} to get to $F(1)=2$, in order for the static solutions to have the same energy density. From this relation, we obtain that $F_Y(Y=1)$ should also be unit, that is, $F_Y(1)=1$.

These are the general restrictions on $F(Y)$ to make the model twin of the standard model. For instance, in the first example studied before, if one writes
\be
F(Y)=\frac{2n-1}{n} \left(1+\frac{Y|Y|^{n-1}}{2n-1}\right)
\ee
one gets to the Lagrange density
\be
{\cal L}=\frac{\partial_\mu \phi \partial^\mu\phi}{2^n \,n}\left|\frac{\partial_\mu \phi \partial^\mu\phi}{U(\phi)}\right|^{n-1} \!\!- \frac{2n-1}{n}U(\phi)
\ee
which is twin of the standard model with the same $U(\phi)$.
Similarly, we can rewrite the model with $F(Y)$ given by Eq.÷\eqref{FTqReg} in the form
\be\label{Taon3}
\LL=-{\sqrt{3}}\,U(\phi)\left(\sqrt{1-\frac{\partial_\mu \phi \partial^\mu \phi}{U(\phi)}}-{\sqrt{3}}\right) 
\ee
with $a$ and $b$ chosen to offer another family of twin models to the standard model with the same $U(\phi)$.

We go on, and we consider the function
\be
F(Y)=1+Y+\frac{\alpha}{3} \left(1-Y\right)^3
\ee
with $\alpha$ real constant. This is the simplest polynomial function which makes the model twin of the standard model, with $F(Y)=1+Y$.
For this choice of $F(Y)$, we note that $A^2=1$, so the two models are twins, and they also have the very same stability features.
This result is interesting since it shows that there are twinlike models that cannot be distinguished by linear stability, differently from the systems investigated in Refs.~\cite{altw,bdglm}.

\section{Comments and conclusions}
\label{end}

In this work we studied the presence of kinks in models described by a single real scalar field in bidimensional spacetime.
We introduced a family of models, which is defined by the two functions $F(Y)$ and $U(\phi)$, and has the general structure
${\cal L}=-U(\phi)F(Y)$, where we used $Y=-{\partial_\mu\phi\partial^\mu\phi}/{2U(\phi)}$,
with $U(\phi)$ being a function of the real scalar field $\phi$. We note that for $F(Y)=1+Y$ the model becomes the standard model,
described by Eq.~\eqref{sm}.

As an interesting result, we introduced the family of models defined by
\be
F(Y)=1+Y+\frac{\alpha}{3}(1-Y)^3\nonumber
\ee
which provides twinlike models to the standard model, with $F(Y)=1+Y$. For such models, the presence of defects and their stability depend on the choice of $U(\phi)$, but for each $U(\phi)$, the standard and twin models have the very same stability features.

The issue concerning twinlike models deserves further investigations, and we are now considering the possibility to extend the idea to the case of more sophisticated defects, such as vortices and monopoles, with the addition of Abelian and non-Abelian gauge fields. Another issue concerns the presence of fermions, leading us to supersymmetric extensions of the twinlike models here introduced.



\begin{references}


\bb{vs}A. Vilenkim and E.P.S. Shellard, Cosmic strings andother topological defects (Cambridge, Cambridge/UK,1994).

\bb{ms}N. Manton and P. Sutcliffe, Topological solitons (Cam-bridge, Cambridge/UK, 2004).\bb{wa} D. Walgraef, Spatio-temporal pattern formation (Springer-Berlag, New York, 1997).

\bb{tt}M. Toharia and M. Trodden, Phys. Rev. Lett.{\bf 100}, 041602 (2008).
\bb{dunne}G. Basar and G.V. Dunne, Phys. Rev. Lett. {\bf100}, 200404 (2008).

\bb{ba}A. Vanhaverbeke, A. Bischof, and R. Allenspach, Phys. Rev. Lett. {\bf101}, 107202 (2008).

\bb{juan}A. Alonso-Izquierdo, M. A. Gonzalez Leon, and J. Mateos Guilarte, Phys. Rev. Lett. {\bf101}, 131602 (2008).

\bb{bec1}J. Belmonte-Beitia, V.M. P«erez-Garcia, V. Vekslerchik,and V.V. Konotop, Phys. Rev. Lett. {\bf100}, 164102 (2008).

\bb{bec2}A.T. Avelar, D. Bazeia, and W.B. Cardoso, Phys. Rev. E {\bf79}, 025602(R) (2009).

\bb{ak}R. Auzzi and S. Prem Kumar, Phys. Rev. Lett. {\bf103}, 231601 (2009). 
\bb{mr}A.D. Martin and J. Ruostekoski, Phys. Rev. Lett. {\bf104}, 194102 (2010).

\bb{exp}Haiyun Liu et al., Phys. Rev. Lett. {\bf105}, 027001 (2010).

\bb{rs}T. Romanczukiewicz and Ya. Shnir, Phys. Rev. Lett. {\bf105}, 081601 (2010).

\bb{as}M. Angeles Perez-Garcia, J. Silk, and J. R. Stone, Phys. Rev. Lett. {\bf105}, 141101 (2010).

\bb{lump}A.T. Avelar, D. Bazeia, L. Losano, R. Menezes, Eur. Phys. J. C {\bf55}, 133 (2008);
A.T. Avelar, D. Bazeia, W.B. Cardoso, and L. Losano, Phys. Lett. A {\bf374}, 222 (2009). 

\bb{c1}C. Armendariz-Picon, T. Damour, and V. F. Mukhanov, Phys. Lett. B {\bf458}, 209 (1999);
J. Garriga and V. F. Mukhanov, Phys. Lett. B {\bf458}, 219 (1999); T. Chiba, T. Okabe, and M. Yamaguchi, Phys. Rev. D {\bf62}, 023511 (2000).

\bb{s1}E. Babichev, Phys. Rev. D {\bf74}, 085004 (2006).
\bb{s2}D. Bazeia, L. Losano, R. Menezes and J.C.R.E. Oliveira, Eur. Phys. J. C {\bf51}, 953 (2007).
\bb{s3}C. Adam, J. Sanchez-Guillen and A. Wereszczynski, J. Phys A {\bf40}, 13625 (2007); {\bf42}, 089801 (E) (2009).
\bb{b1}M. Olechowski, Phys. Rev. D {\bf78}, 084036 (2008).
\bb{b2}C. Adam, N. Grandi, P. Klimas, J. Sanchez-Guillen and
A. Wereszczynski, J. Phys. A {\bf41}, 212004 (2008).
\bb{b3}D. Bazeia, L. Losano and R. Menezes, Phys. Lett. B {\bf668},
246 (2008).
\bb{b4}D. Bazeia, A. R. Gomes, L. Losano and R. Menezes, Phys. Lett. B {\bf671}, 402 (2009).
\bibitem{acg}N.~Afshordi, D.J.H.~Chung and G.~Geshnizjani, Phys.\ Rev.\  D {\bf 75}, 083513 (2007).

\bb{B}H. T. Li, Y. X. Liu, Z. H. Zhao and H. Guo, Phys. Rev. D {\bf83}, 045006 (2011); R.A.C. Correa, A. de Souza Dutra, and M.B. Hott, Class. Quant. Grav. {\bf28}, 155012 (2011); R.C. Fonseca, F.A. Brito, and L. Losano, Phys. Lett. B {\bf697}, 493 (2011); W.T. Cruz, A.R. Gomes, and C.A.S. Almeida, EPL {\bf96}, 31001 (2011); R.R. Landim, G. Alencar, M.O. Tahim, and N.R. Costa Filho, JHEP {\bf1108}, 071 (2011); D. Bazeia, F. A. Brito, F. G. Costa, Phys. Lett. B {\bf704}, 631 (2011).

\bibitem{altw} M. Andrews, M. Lewandowski, M. Trodden, and D. Wesley, Phys.\ Rev.  D {\bf82}, 105006 (2010).
\bb{bdglm}D. Bazeia, J.D. Dantas, A.R. Gomes, L. Losano, and R. Menezes, Phys. Rev. D {\bf84}, 045010 (2011).
\bb{aq}C. Adam and J.M. Queiruga, {\it An algebraic contruction of twin-like models}. [arXiv:1109.4159].

\bibitem{sen}
  A.~Sen, Phys.\ Rev.\  D {\bf68}, 066008 (2003); Int.\ J.\ Mod.\ Phys.\ A {\bf20}, 5513 (2005).
\bb{bmr}D. Bazeia, R. Menezes, and J.G. Ramos, Mod. Phys. Lett. A {\bf20}, 467 (2005).
\end{references}
\end{document}